\documentclass[prl,twocolumn]{revtex4-1}
\usepackage{times}
\usepackage{amsmath}
\usepackage{amsfonts}
\usepackage{graphicx}
\usepackage{hyperref}
\usepackage{dcolumn}
\usepackage{bm}

\def\beq{\begin{equation}}
\def\eeq{\end{equation}}
\def\bea{\begin{eqnarray}}
\def\eea{\end{eqnarray}}
\def\beqn{\begin{eqnarray}} 
\def\eeqn{\end{eqnarray}}
\def\beeq{\begin{eqnarray}}
\def\eeeq{\end{eqnarray}}

\def\nn{\nonumber}
\def\Eq#1{Eq.~(\ref{#1})}

\def\qon#1{q_{#1,0}^{(+)}}

\def\qb{\mathbf{q}}

\def\lb{\boldsymbol{\ell}}

\def\uv{{\rm UV}}

\def\gs{g_{\rm S}}

\def\v{{\rm V}}
\def\r{{\rm R}}

\def\ii{\imath 0}

\def\ad#1{{\cal A}_{\rm D}^{(#1)}}
\def\af#1{{\cal A}_{\rm F}^{(#1)}}

\def\aduvir#1{{\cal A}_{\uv/{\rm C}}^{(#1)}}

\def\ps#1{\widetilde \Delta_{#1}}

\usepackage{xcolor}

\definecolor{darkorange}{rgb}{1.0, 0.55, 0.0}

\begin{document}


\title{Rewording Theoretical Predictions at Colliders with Vacuum Amplitudes}

\author{Selomit Ram\'{\i}rez-Uribe~$^{(a,b)}$} \email{selomitru@uas.edu.mx} 
\author{Prasanna K. Dhani~$^{(a)}$} \email{dhani@ific.uv.es}
\author{German F.R. Sborlini~$^{(c)}$} \email{german.sborlini@usal.es} 
\author{Germ\'an Rodrigo~$^{(a)}$}\email{german.rodrigo@csic.es}
\affiliation{${}^{a}$ Instituto de F\'{\i}sica Corpuscular, Universitat de Val\`{e}ncia -- Consejo Superior de Investigaciones Cient\'{\i}ficas, Parc Cient\'{\i}fic, E-46980 Paterna, Valencia, Spain. \\
${}^{b}$ Facultad de Ciencias F\'{\i}sico-Matem\'aticas, Universidad Aut\'onoma de Sinaloa, Ciudad Universitaria, CP 80000 Culiac\'an, Mexico. \\
${}^{c}$ Departamento de F\'isica Fundamental e IUFFyM, Universidad de Salamanca, 37008 Salamanca, Spain.}

\date{July 5, 2024}

\begin{abstract}
We propose multiloop vacuum amplitudes in the loop-tree duality (LTD) as the optimal building blocks for efficiently assembling theoretical predictions at high-energy colliders. This hypothesis is strongly supported by the manifestly causal properties of the LTD representation of a vacuum amplitude. The vacuum amplitude in LTD, acting as a kernel, encodes all the final states contributing to a given scattering or decay process through residues in the on-shell energies of the internal propagators. Gauge invariance and the wave function renormalisation of the external legs are naturally incorporated. This methodological approach, dubbed LTD causal unitary, leads to a novel differential representation of cross sections and decay rates that is locally free of ultraviolet and infrared singularities at all orders in perturbation theory. Threshold singularities also match between different phase-space residues. Most notably, it allows us to conjecture for the first time the local functional form of initial-state collinear splitting functions. The fulfillment of all these properties provides a theoretical description of differential observables at colliders that is well defined in the four physical dimensions of the space-time. 
\end{abstract}

\maketitle


More than 10 years after the discovery of the Higgs boson~\cite{ATLAS:2012yve,CMS:2012qbp}, the current Run 3 of the CERN's Large Hadron Collider~(LHC), the upcoming high luminosity phase and the planned  experiments~\cite{FCC:2018byv,ILC:2013jhg,Linssen:2012hp,CEPCStudyGroup:2018ghi,Long:2020wfp}, make the need for better theoretical predictions at high-energy colliders more pressing than ever~\cite{Heinrich:2020ybq}. In most cases, the theoretical uncertainty is the main limiting factor in achieving accurate measurements and thereby discern potential signals of new physics~\cite{EuropeanStrategyGroup:2020pow,Butler:2023glv}.

Scattering amplitudes at high perturbative orders in quantum field theory are the central objects that allow us to derive precise theoretical predictions for scattering and decay processes. They exhibit very interesting mathematical properties and geometric interpretations~\cite{Arkani-Hamed:2013jha}, spanning the interest in their study beyond collider physics, e.g., in gravitational physics~\cite{Cheung:2018wkq,Bern:2019nnu,Foffa:2019hrb,Blumlein:2021txj}. Highly efficient and sophisticated methods have been developed~\cite{Abreu:2022mfk}.

Scattering amplitudes, however, are defined for a fixed number of external particles, and different scattering amplitudes squared with different numbers of external particles must be combined to obtain theoretical predictions. Fixing the number of external particles in theories like Quantum Chromodynamics, where particles can be emitted at exactly zero energies or collinear to each other, makes scattering amplitudes tricky, particularly in four space-time dimensions. In addition, loop configurations involve extreme quantum fluctuations at infinite energy, where the theory is no longer valid. As a consequence, scattering amplitudes are evaluated in arbitrary space-time dimensions, e.g. by dimensional regularization (DREG)~\cite{Bollini:1972ui,tHooft:1972tcz}, where these ambiguities from infrared~(IR) and ultraviolet~(UV) configurations translate into poles of the extra dimensions.  

In this Letter, we present a novel methodology to compute differential cross sections and decay rates at high perturbative orders directly in the four physical dimensions of the space-time. It is inspired by the community efforts to circumvent the bottleneck imposed by working in arbitrary space-time dimensions~\cite{TorresBobadilla:2020ekr,Gnendiger:2017pys,Pittau:2012zd,Pereira:2022mne,Gnendiger:2017rfh,Fazio:2014xea,Soper:1998ye,Soper:1999xk}. The method is based on the loop-tree duality~(LTD)~\cite{Catani:2008xa,Bierenbaum:2010cy,Bierenbaum:2012th,Buchta:2014dfa,Buchta:2015wna,Tomboulis:2017rvd,Driencourt-Mangin:2017gop,Jurado:2017xut,Driencourt-Mangin:2019aix,Runkel:2019yrs,Baumeister:2019rmh,Aguilera-Verdugo:2019kbz,Runkel:2019zbm,Capatti:2019ypt,Driencourt-Mangin:2019yhu,Capatti:2019edf,Aguilera-Verdugo:2020set,snowmass2020,Plenter:2020lop,Plenter:2019jyj,Aguilera-Verdugo:2020kzc,Ramirez-Uribe:2020hes,JesusAguilera-Verdugo:2020fsn,Sborlini:2021owe,TorresBobadilla:2021ivx,Bobadilla:2021pvr,deJesusAguilera-Verdugo:2021mvg,Benincasa:2021qcb,Kromin:2022txz,Ramirez-Uribe:2021ubp,Clemente:2022nll,Ramirez-Uribe:2024wua,EPpatent,deLejarza:2024pgk,Rios-Sanchez:2024xtv,Hernandez-Pinto:2015ysa,Sborlini:2016gbr,Sborlini:2016hat,Prisco:2020kyb}, which transforms multiloop Feynman amplitudes into tree amplitudes by integrating out one degree of freedom of each loop momenta through the Cauchy's residue theorem~(CRT). 
Complementary approaches include~\cite{Agui-Salcedo:2023wlq,Borinsky:2022msp,Sterman:2023xdj}. Our proposal generalizes the local method of four-dimensional unsubtraction~(FDU)~\cite{Hernandez-Pinto:2015ysa,Sborlini:2016gbr,Sborlini:2016hat,Prisco:2020kyb}, 
while introducing relevant improvements and new features that facilitate an efficient implementation. 

{\it Starting hypothesis.} -- Since fixing the number of external particles is at the origin of many of the difficulties in obtaining theoretical predictions from scattering amplitudes, we propose a radical solution which consists in eliminating all external particles. Our starting hypothesis is that all quantum fluctuations contributing to a scattering or a decay process at high energies are contained in vacuum amplitudes, i.e., in scattering amplitudes without external particles. Then, by exploiting the manifestly causal behaviour of the LTD representation and its connection with directed acyclic graphs~(DAGs)~\cite{Ramirez-Uribe:2021ubp,Clemente:2022nll,Ramirez-Uribe:2024wua}, we generate all final states from residues on causal propagators, \Eq{eq:causalpropagator}, of the vacuum amplitude. At first glance, there is a clear advantage: the same vacuum amplitude could act as the kernel of different scattering and decay processes. Moreover, it also avoids many of the ambiguities that arise in attempting to construct loop amplitudes from the forward limit of tree-level amplitudes~\cite{Caron-Huot:2010fvq,He:2015yua,Runkel:2019zbm}, while allowing us to benefit from the know-how on multiloop scattering amplitudes. The method is manifestly gauge invariant and naturally incorporates the wave-function renormalisation of the external legs. 

The integrand of a vacuum amplitude is a scalar and depends on $\Lambda$ independent loop momenta, $\{\ell_j\}_{j=1,\ldots, \Lambda}$, 
\beq
{\cal A}^{(\Lambda)}= \int_{\ell_1\cdots \ell_\Lambda} \af{\Lambda}~, \qquad \int_{\ell_j} = -\imath \mu^{4-d} \int \frac{d^d\ell_j}{(2\pi)^d}~,
\label{eq:feynman}
\eeq
in $d$ space-time dimensions. The scale $\mu$ represents the DREG scale. Although not necessary, we keep the dimensional dependence to define the method in a general manner and to confront intermediate results with DREG expressions. The subindex ${\rm F}$ stands for the Feynman representation, which is written in terms of Feynman propagators, $G_F(q_{i_s}) = (q_{i_s}^2-m_{i_s}^2+\ii)^{-1}$ and a numerator that depends on the theory under consideration. The momenta of these propagators are denoted by $q_{i_s}$, where $q_{i_s}$ are linear combinations of the independent loop momenta, and $m_{i_s}$ are their masses. The LTD representation is obtained from the Feynman representation in \Eq{eq:feynman} by integrating out one degree of freedom per loop via CRT. If the energy components are integrated out, then the vacuum amplitude in LTD has support in the Euclidean space of the spatial components of the loop momenta~\footnote{For convenience, the loop integration measure differs from the integration measure in Ref.~\cite{Aguilera-Verdugo:2020set} by a factor, which is reabsorbed in $\ad{\Lambda}$},
\beq
{\cal A}^{(\Lambda)}= \int_{\lb_1\cdots \lb_\Lambda} \ad{\Lambda}~, \qquad \int_{\lb_j} = \mu^{4-d} \int \frac{d^{d-1}\lb_j}{(2\pi)^{d-1}}~.
\label{eq:LTD}
\eeq
The Euclidean nature of the integration domain has some advantages, e.g. for analytic asymptotic expansions~\cite{Plenter:2020lop,Plenter:2019jyj,Driencourt-Mangin:2017gop}. In Eq.~(\ref{eq:LTD}), $\ad{\Lambda}$ is the dual vacuum amplitude, i.e. the integrand of the vacuum amplitude in LTD. 

It is highly non-trivial that only causal singularities remain in LTD after summation of all nested residues obtained by the recursive application of CRT, while potential noncausal or unphysical singularities are analytically cancelled out. This property has important consequences for the numerical stability of the integrand~\cite{Ramirez-Uribe:2020hes}. The cancellation of noncausal singularities was observed first at one loop~\cite{Buchta:2014dfa,Buchta:2015wna,Capatti:2019edf} and then at two loops~\cite{Aguilera-Verdugo:2019kbz}, while the existence of a manifestly causal representation was conjectured to hold to all loop orders in~\cite{Aguilera-Verdugo:2020set,snowmass2020}, and has been further investigated in~\cite{Aguilera-Verdugo:2020kzc, Ramirez-Uribe:2020hes,JesusAguilera-Verdugo:2020fsn,Sborlini:2021owe,TorresBobadilla:2021ivx,Bobadilla:2021pvr,deJesusAguilera-Verdugo:2021mvg,Benincasa:2021qcb,Kromin:2022txz,Ramirez-Uribe:2021ubp,Clemente:2022nll,Ramirez-Uribe:2024wua,deLejarza:2024pgk,Rios-Sanchez:2024xtv,EPpatent}. Causal singularities are those that can be interpreted as arising from kinematic configurations in which all particles crossing a line that divides an amplitude into two subamplitudes simultaneously become on-shell states or real particles with their momenta aligned in the same direction. They are responsible for the absorptive (imaginary) part of the amplitude due to thresholds and for the emergence of IR  singularities, which appear as pinched thresholds for massless particles. In addition, LTD provides a clear physical interpretation of overlapping discontinuities in scattering amplitudes through the concordance of compatible causal singularities in a simpler way than the Landau~\cite{Landau:1959fi} and Steinmann~\cite{steinmann1,steinmann2} relations.
 
Because we are considering a vacuum amplitude in LTD, all the Feynman propagators in the Feynman representation are substituted by causal propagators of the form~\cite{Aguilera-Verdugo:2020set}
\beq
\frac{1}{\lambda_{i_1 i_2 \cdots i_n}} = \frac{1}{\sum_{s=1}^n \qon{i_s}}~,
\label{eq:causalpropagator}
\eeq
with $\qon{i_s} = \sqrt{\qb_{i_s}^2+ m_{i_s}^2-\ii}$ the on-shell energies of the internal momenta, where $\qb_{i_s}$ are the spatial components. The factor $\ii$ stems from the original infinitesimal complex prescription of the Feynman propagators. Tracing causality with a Lorentz covariant complex formulation was one of the main results of the LTD's seminal works~\cite{Catani:2008xa,Bierenbaum:2010cy,Bierenbaum:2012th}. Terms of the form~(\ref{eq:causalpropagator}) are called causal propagators because they involve a set of internal particles that divide the vacuum amplitude into two subamplitudes, with the momentum flow of all particles in the set aligned in the same direction. Each term in a dual vacuum amplitude is proportional to a product of causal propagators in which the momenta of the shared particles are aligned. The numerators of a vacuum amplitude in LTD are generically written in terms of on-shell energies and  internal masses. 

The on-shell energies are  by definition positive in the limit of a vanishing complex prescription. Therefore, the dual vacuum amplitude becomes singular at $\lambda_{i_1 i_2 \cdots i_n} \to 0$ only in the soft limits where several on-shell energies vanish simultaneously, $\qon{i_s}=0~\forall~s \in \{1, \ldots, n\}$. This is actually not a soft singularity because it is compensated by the low energy behaviour of the integration measure~\footnote{Feynman propagators raised to a power in scalar theories could modify this dimensional argument. In QCD/QED, the numerator of the vacuum amplitude restores this dimensional analysis~\cite{LTD:2024yrb}}. Consequently, aside of being unaffected by noncausal singularities, the dual vacuum amplitude is also free of causal thresholds and IR singularities. This is consistent with the absence of external particles~\cite{Aguilera-Verdugo:2020set}. Therefore, the integrated dual vacuum amplitude is purely real, up to a global phase.

The central idea of our proposal is that by analytically continuing the on-shell energies of the particles that will be identified as incoming to negative values, and by considering different residues on the causal propagators, $1/\lambda_{i_1 i_2 \cdots i_n}$, dubbed phase-space residues, all the amplitude interferences with different numbers of external particles contributing to a differential cross section or a decay rate are generated at once, with the dual vacuum amplitude acting as a kernel for all contributions. Since all contributions are generated from the same vacuum amplitude, collinear, soft and threshold singularities match locally among the different phase-space residues, with the sole exception of initial-state collinear singularities whose local cancellation is limited by kinematics. Note that the vacuum amplitude includes selfenergy insertions. Therefore, the wave-function renormalisation of the external legs is naturally incorporated. In addition, the vacuum amplitude is gauge invariant, so the phase-space residues are also gauge invariant. For the sake of the reader, a detailed calculation of decay processes is presented in~\cite{LTD:2024yrb}.


{\it Vacuum amplitudes and LTD causal unitary. --} We propose the following differential representation of the N$^k$LO contribution to a physical observable
\bea
\label{eq:mastercausal}
&& d\sigma_{{\rm N}^k{\rm LO}} = \frac{d\Lambda}{4\sqrt{(p_a\cdot p_b)^2-(m_a m_b)^2}} \,  \\ && \times 
\sum_{(i_1\cdots i_n a b) \in \Sigma} \ad{\Lambda, {\rm R}}(i_1\cdots i_n a b) \, 
{\cal O}_{i_1\cdots i_n} \, \ps{i_1\cdots i_n \bar a \bar b}~, \nn 
\eea
which is obtained as the sum of UV renormalized and initial-state-collinear subtracted phase-space residues
\bea
\ad{\Lambda,\r}(i_1 \cdots i_n a b) &=& {\rm Res} \left(\frac{x_{ab}}{2} \, \ad{\Lambda}, \lambda_{i_1\cdots i_n a b}\right) \nn \\
&-& \aduvir{\Lambda}(i_1 \cdots i_n a b)~,
\label{eq:adresidues}
\eea
where the first term on the r.h.s of \Eq{eq:adresidues} is the residue of the dual vacuum amplitude $\ad{\Lambda}$ at $\lambda_{i_1\cdots i_n a b} = 0$. The number of loops of $\ad{\Lambda}$ is $\Lambda = L+N-1$, where $N$ is the total number of external particles in leading order (LO) kinematics, and~$L$ is the maximum number of loops that contribute at N$^k$LO. Typically, $L=2$ at next-to-next-to-leading order~(NNLO). Each residue on the dual vacuum amplitude implements a different $n$-particle final state, and $\Sigma$ is the set of all final states contributing at N$^k$LO, with $n\in \{m, \ldots, m+k\}$, where $m$ is the number of final-state particles in LO kinematics, i.e., $m=N-2$ for a scattering process.

The indices of the initial-state particles are $a$ and $b$, and their on-shell energies are analytically continued to negative values. We  define $x_{ab} = 4 \qon{a}\qon{b}$. The master differential representation introduced in~\Eq{eq:mastercausal} is also valid for decay processes, where $m=N-1$, provided that the initial-state flux $4\sqrt{(p_a\cdot p_b)^2-(m_a m_b)^2}$ is replaced by $1/(2 m_a)$ and $x_{ab}\to x_a = 2 \qon{a}$. 

The counterterm $\aduvir{\Lambda}$ implements a local UV renormalisation, and a local subtraction of initial-state collinear singularities. Here, we  introduce a novelty with respect to previous implementations of a local UV renormalisation~\cite{Driencourt-Mangin:2019aix} (see also~\cite{Becker:2010ng,Donati:2013voa,Cherchiglia:2020iug,Rios-Sanchez:2024xtv}). We extract the leading UV behaviour directly in LTD by rescaling the on-shell energies:
\beq
\qon{i_s} \to \sqrt{(\rho \, \lb_{s} + \lb_r)^2+m_{i_s}^2 + (\rho^2-1) \mu_{\uv}^2-\ii}~,
\label{eq:uvexpansion}
\eeq
and then by expanding the phase-space residues for $\rho\to \infty$. In \Eq{eq:uvexpansion}, we assume $\qb_{i_s} = \lb_s + \lb_r$, with $\lb_s$ in the UV region, while $|\lb_r| \ll |\lb_s|$. The scale $\mu_\uv$ acts as the renormalisation scale, and subleading UV terms are added to fix the renormalisation scheme~\cite{LTD:2024yrb}. 

The integration measure for a scattering process is 
\beq
d\Lambda = \prod_{j=1}^{\Lambda-2} d\Phi_{\lb_j} = \prod_{j=1}^{\Lambda-2} \mu^{4-d} \frac{d^{d-1} \lb_j}{(2\pi)^{d-1}}~, 
\label{eq:integrationmeasure}
\eeq
which is written in terms of the spatial components of $\Lambda-2$ primitive loop momenta. Two of the loop three-momenta are fixed by the initial state.  The Dirac-delta function in the spatial components of the external momenta is absent because momentum conservation in these components is self-satisfied in the vacuum amplitude. Energy conservation is imposed by the Dirac-delta function 
\beq
\widetilde \Delta_{i_1\cdots i_n \bar a\bar b} = 2\pi \,  \delta(\lambda_{i_1\cdots i_n \bar a\bar b})~, 
\label{eq:deltaenergy}
\eeq
where
\beq
\lambda_{i_1\cdots i_n \bar a \bar b} = \sum_{s=1}^n \qon{i_s} - \qon{a} - \qon{b}~.
\eeq
The bar over $a$ and $b$ indicates that the corresponding on-shell energies bear a minus sign. 

Finally, the function ${\cal O}_{i_1\cdots i_n}$ encodes the observable under consideration by 
mapping the spatial components of the internal momenta of the dual vacuum amplitude onto the spatial components of the momenta of the external particles. The external on-shell energies are set by the corresponding equation of motion. The default choice ${\cal O}_{i_1\cdots i_n}=1$ gives the total cross section or decay rate after integration.

The dual vacuum amplitude is free of IR and threshold singularities because on-shell energies are positive, as discussed above. However, the phase-space residues exhibit the expected singularities because some of the on-shell energies are promoted to negative values. We show next how collinear, soft and threshold singularities match out locally between phase-space residues. Note that in the spirit of FDU~\cite{Hernandez-Pinto:2015ysa,Sborlini:2016gbr,Sborlini:2016hat,Prisco:2020kyb} and unlike the Soper's~\cite{Soper:1998ye,Soper:1999xk} and local unitarity~\cite{Capatti:2020xjc,Capatti:2022tit} approaches, we do not smear or rescale the momenta to eliminate the phase-space integration constraints. Instead, we work with the actual momenta of the external particles, which is more convenient for predicting differential observables.


{\it Collinear, soft and threshold singularities in the final state. --} Consider three particles, $i_1$, $i_2$ and $i_3$, interacting through a trivalent vertex as an insertion in a vacuum amplitude. The interaction vertex could include loops or effective operators. The dual vacuum amplitude contains a term~\cite{Aguilera-Verdugo:2020kzc,Ramirez-Uribe:2020hes,JesusAguilera-Verdugo:2020fsn,Ramirez-Uribe:2021ubp} proportional to 
\beq
\ad{\Lambda} \sim \frac{1}{\lambda_{i_1 i_2 \cdots ab} \lambda_{i_3 \cdots ab}}~,
\label{eq:doslambdas}
\eeq
where the dots represent other particles acting as  spectators. Here, $a$ and $b$ are identified with the initial-state particles and their on-shell energies are promoted to negative values. 

\begin{figure}[t]
\begin{center}
\includegraphics[scale=0.6]{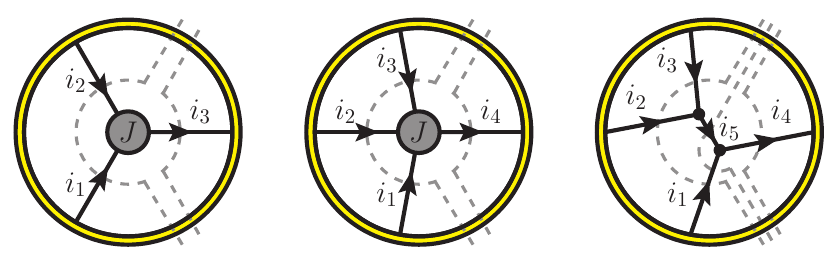}
\caption{Diagrammatic interpretation of the local cancellation of final-state double-collinear (left), and triple-collinear (middle) singularities, including quasicollinear configurations. Phase-space residues represented as dashed lines. The blob $J$ stands for a multiloop sub-diagram, a contact interaction or a combination of trivalent interactions (e.g. right). Each phase-space residue represents an amplitude interference with different numbers of external particles or loops. 
\label{fig:finalstate}}
\end{center}
\end{figure}

The phase-space residues of $\ad{\Lambda}$ are such that $i_3$ becomes on shell for a kinematic configuration with $m$ external particles, while $i_1$ and $i_2$ become on-shell for a configuration with $m+1$. For massless particles and in the limit $\lambda_{i_1 i_2 \bar i_3} = \qon{i_1} + \qon{i_2} - \qon{i_3} \to 0$, each of these contributions develops a collinear singularity, but the sum of both is finite:
\bea
 \lim_{\lambda_{i_1 i_2 \bar i_3}\to 0} &&\left( \ad{\Lambda}(i_1 i_2 \cdots a b) \, \ps{i_1 i_2 \cdots \bar a \bar b} \right. \nn \\ &&
\left. + \ad{\Lambda}{(i_3 \cdots a b)} \, \ps{i_3 \cdots \bar a \bar b}\right) = {\cal O} (\lambda_{i_1 i_2 \bar i_3}^0)~.
\label{eq:doublecollinear}
\eea
A diagrammatic interpretation is shown in Fig.~\ref{fig:finalstate}. This is a schematic interpretation in terms of causal propagators, which may correspond to the sum of several Feynman diagrams. Each of the phase-space residues generates a potential singularity at $\lambda_{i_1 i_2 \bar i_3} \to 0$, although with opposite signs, 
\bea
&& \left.  \frac{1}{\lambda_{i_1i_2\cdots ab}} \right|_{\lambda_{i_3\cdots ab}=0} = \frac{1}{\lambda_{i_1i_2\bar i_3}}~, \nn \\ 
&& \left.  \frac{1}{\lambda_{i_3\cdots ab}} \right|_{\lambda_{i_1i_2\cdots ab}=0} = -\frac{1}{\lambda_{i_1i_2\bar i_3}}~. 
\label{eq:localdouble}
\eea
The cancellation of this collinear singularity is not restricted by the phase space:
\beq
\lim_{\lambda_{i_1 i_2 \bar i_3}\to 0} \ps{i_1 i_2 \cdots \bar a \bar b} = \ps{i_3 \cdots \bar a \bar b}~, 
\eeq
and is therefore fully local. The factors accompanying the causal propagators in~\Eq{eq:doslambdas} in the full vacuum amplitude also match in the collinear limits of the phase-space residues, thus ensuring a local cancellation. For diagrams with selfenergy insertions, the phase-space residues behave individually as $1/\lambda_{i_1i_2\bar i_3}^2$. These quadratic singularities also cancel out, along with the associated subleading linear singularities (\cite{LTD:2024yrb} and Supplemental Material). 

It is remarkable that a similar behaviour holds for quasicollinear configurations with massive particles because particle masses appear implicitly in the causal propagators through the on-shell energies. Although collinear singularities are shadowed in the form of large mass logarithms, these logarithms also cancel when $\lambda_{i_1 i_2 \bar i_3} \lesssim {\cal O}(m_{i_s})$.  This fact provides a seamless transition between a massive and a massless calculation, allowing for an almost identical implementation~\cite{Sborlini:2016hat}.

Likewise, the local cancellation of triple-collinear singularities is described by the sum of two phase-space residues, one of them involving the three particles that become collinear and the other concerning the parent parton 
\bea
&&\lim_{\lambda_{i_1 i_2 i_3 \bar i_4}\to 0} \left( \ad{\Lambda}(i_1 i_2 i_3 \cdots a b) \, \ps{i_1 i_2 i_3 \cdots \bar a\bar b} \right. \nn \\
&& \qquad \left. + \ad{\Lambda}{(i_4 \cdots a b)} \, \ps{i_4 \cdots \bar a\bar b} \right) 
= {\cal O} (\lambda_{i_1 i_2 i_3\bar i_4}^0)~. 
\eea
The proof is very similar to that in \Eq{eq:localdouble}. Moreover, it is independent of the internal structure of the quadrivalent interaction, which is represented by the blob $J$ in Fig.~\ref{fig:finalstate}. If we now consider the specific collinear sequence $i_4 \to i_1 i_5 (i_5\to i_2 i_3)$, with $\lambda_{i_2 i_3 \bar i_5}$ going faster to zero than $\lambda_{i_1 i_2 i_3 \bar i_4}$, we may explore in detail the overlapping singularities through the sum of the three phase-space residues $\ad{\Lambda}{(i_1 i_2 i_3 \cdots a b)}$, $\ad{\Lambda}{(i_1 i_5 \cdots a b)}$ and $\ad{\Lambda}{(i_4 \cdots a b)}$. The proof that their sum is not singular in either of the collinear limits is a little trickier, and involves the three permutations of
\beq
\left.  \frac{1}{\lambda_{i_1i_2i_3\cdots ab}\lambda_{i_1i_5\cdots ab}} \right|_{\lambda_{i_4\cdots ab}=0} 
= \frac{1}{\lambda_{i_1i_2i_3\bar i_4}\lambda_{i_1i_5\bar i_4}}~.
\eeq

\begin{figure}[t]
\begin{center}
\includegraphics[scale=0.6]{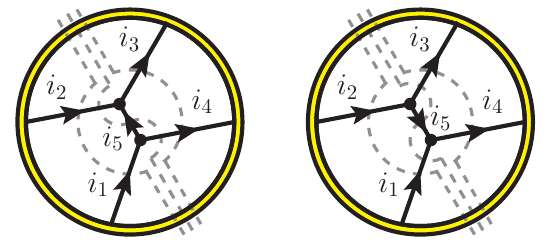}
\caption{Phase-space residues (dashed lines) contributing to final-state soft and threshold singularities. 
\label{fig:soft}}
\end{center}
\end{figure}

For the case of soft and threshold singularities, we need to analyse terms proportional to  
\beq
\ad{\Lambda} \sim \frac{1}{\lambda_{i_1i_2 \cdots ab}} 
\left( \frac{1}{\lambda_{i_2i_4 i_5 \cdots ab}} + \frac{1}{\lambda_{i_1i_3i_5 \cdots ab}} \right)\frac{1}{\lambda_{i_3i_4 \cdots ab}}~.
\label{eq:soft}
\eeq
A graphical interpretation is presented in Fig.~\ref{fig:soft}. Soft singularities occur at the common endpoint of two collinear configurations. The momentum of the soft particle, denoted~$i_5$, flows in opposite directions in each of them. The two configurations become compatible exactly at $\qon{i_5}=0$, where the momentum direction of $i_5$ does not matter. The local cancellation of the soft singularity is ensured because the sum of the residues from each of the four causal propagators in \Eq{eq:soft} match each other in the soft limit. 

On the other hand, thresholds in loop contributions occur whenever internal particles are produced as external on-shell particles. These configurations are integrable and responsible for the imaginary component of the loop amplitude. The phase-space residues may exhibit threshold singularities and thus an imaginary component whereas the dual vacuum amplitude is real. One might expect that the imaginary components, and therefore threshold singularities, could match in the sum of phase-space residues. Specifically, for the terms considered in~\Eq{eq:soft}
\bea
&& {\rm Res} \left(\ad{\Lambda}(i_1 i_2 \cdots ab) \, \ps{i_1 i_2 \cdots \bar a\bar b}  \right.  \\
&& \qquad \left. + \ad{\Lambda}(i_3 i_4 \cdots ab) \, \ps{i_3 i_4 \cdots \bar a\bar b} , \lambda_{i_1i_2\bar i_3 \bar i_4} 
\right) = 0~, \nn 
\eea
which represents the scattering subprocess $i_1i_2\to i_3i_4$.

Note that the local cancellation of soft and collinear singularities involves phase-space residues with different numbers of external particles, while the local matching of thresholds requires phase-space residues with the same number of external particles and is not restricted by kinematical cuts. The generalisation to arbitrary multi-collinear, soft and threshold configurations proceeds in a similar algorithmic way. 


{\it Collinear, soft and threshold singularities in the initial state. --} Soft and threshold singularities in the initial state also match locally. With respect to final-state configurations, we must consider a different combination of causal propagators, 
\begin{align}
\ad{\Lambda} &\sim \frac{1}{\lambda_{i_1i_2 i_3 i_4}} 
\left( \frac{1}{\lambda_{i_2i_3 i_5}} + \frac{1}{\lambda_{i_1i_4i_5}} \right)\left( \frac{1}{\lambda_{i_1i_2 \cdots}} + \frac{1}{\lambda_{i_3i_4\cdots}} \right) \nn \\ 
&+ \frac{1}{\lambda_{i_2i_3 i_5} \lambda_{i_1i_4i_5}}
\left( \frac{1}{\lambda_{i_2i_4 i_5\cdots}} + \frac{1}{\lambda_{i_1i_3 i_5 \cdots}} \right)~. 
\label{eq:softinitial}
\end{align}
Particles labeled either $(i_1,i_2)$ or $(i_3,i_4)$ are now incoming, and their momentum flows are reversed over the phase-space residues. As for the final state, the matching of thresholds occurs among phase-space residues with the same number of final-state particles, and is thus not limited by kinematics. Soft singularities cancel out locally because the virtual and real phase spaces meet in the soft limit. 

\begin{figure}[t]
\begin{center}
\includegraphics[scale=0.6]{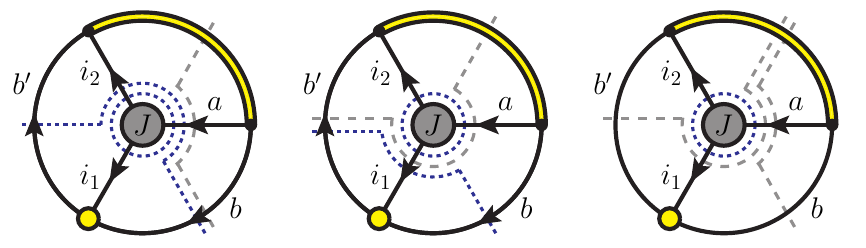}
\caption{Phase-space residues (gray dashed lines) contributing to the initial-state collinear splitting $a\to i_1 i_2$. The initial-state particles $b$ and $b'$ are assumed to be of the same flavor. The dotted blue lines represent causal propagators.
\label{fig:initialstate}}
\end{center}
\end{figure}

Initial-state collinear singularities are a case apart, because their local cancellation is limited by the phase space. The collinear splitting $a\to i_1 i_2$ requires the participation of two incoming spectators, $b$ and $b'$, assuming they are of the same flavour to generate the same initial state. The terms contributing to this collinear splitting are
\bea
\ad{\Lambda} &\sim& \frac{1}{\lambda_{i_1i_2 a}} 
\bigg( 
\frac{1}{\lambda_{i_1\cdots  a b'}} 
\left( \frac{1}{\lambda_{\cdots a b}} 
+ \frac{1}{\lambda_{i_1 b b'} }  \right) \nn \\ &&
+ \frac{1}{\lambda_{i_2 a b b'} \lambda_{\cdots a b}} \bigg)~,
\label{eq:softinitial}
\eea
and have been represented in Fig.~\ref{fig:initialstate}. In the collinear limit the sum of phase-space residues behaves as 
\bea
&& \lim_{\lambda_{i_1 i_2 \bar a}\to 0} \left( {\cal A}_D^{(\Lambda)}(i_1 \cdots a b') \, \ps{i_1 \cdots \bar a\bar b'} \right.  \left. + {\cal A}_D^{(\Lambda)}{(\cdots a b)} \, \ps{\cdots \bar a\bar b} \right) \nn \\
&& \qquad \sim  f(\lambda_{i_1 i_2 \bar a})
 \left( \ps{i_1 \cdots \bar a\bar b'} - \ps{\cdots \bar a\bar b} \right) 
+ {\cal O} (\lambda_{i_1 i_2 \bar a}^0)~,
\label{eq:initialcollinear}
\eea
where $f(\lambda_{i_1 i_2 \bar a})$ is the same singular function for both contributions, but evaluated at different values of the on-shell energies. In the soft limit, $\qon{i_1}\to 0$, both phase spaces coincide and suppress the singular behavior encoded by $f(\lambda_{i_1 i_2 \bar a})$. 

From \Eq{eq:initialcollinear}, another relevant result of this work is that in the collinear limit the functional form encoded by $f(\lambda_{i_1 i_2 \bar a})$ is the same for the loop and tree-level phase-space residues, and hence the unintegrated contributions to the Altarelli-Parisi splitting functions~\cite{Altarelli:1977zs}. For example, the collinear splitting $q\to qg$ is described by the local splitting function
\beq
{\cal P}_{qq}^{(0)}(z_{\v},z;\epsilon) = {\cal P}_{qq}^{(0)}(z_{\v}; \epsilon) 
\left( \delta(z_{\v}-z) - \delta(1-z) \right)~, 
\eeq
where 
\beq
{\cal P}_{qq}^{(0)}(z; \epsilon) = \frac{1+z^2}{1-z} - \epsilon (1-z)
\eeq
is the customary bare splitting function, with $1-z$ the longitudinal momentum fraction of the gluon when emitted as real, and $1-z_{\v}$ the longitudinal momentum fraction of the same gluon in the loop contribution. Integrating over $z_{\v}$, with $z_{\v}\in [0,1]$, we obtain the expected loop factor $3/2 \, \delta(1-z)$.

{\it Discussion. --} We have presented a novel representation of differential observables for high-energy colliders, where all final states contributing to a scattering or a decay process are coherently generated from a multiloop vacuum amplitude in LTD under a common integration domain. Exploiting the manifestly causal properties of vacuum amplitudes in LTD provides a consistent theoretical framework in which certain mathematical artefacts are absent and many technical difficulties are solved in one go. For example, our formalism deals directly with the actual momenta of the external particles, which is more convenient for predicting differential observables. The absence of collinear, soft and threshold singularities in the dual vacuum amplitude leads to a local matching of these singularities in the sum over all phase-space residues to all perturbative orders. Large logarithmic terms from quasicollinear configurations of massive particles are cancelled. Furthermore, the vacuum amplitude consistently incorporates gauge invariance and the wave function renormalisation of external particles, which is crucial for the IR local cancellation with tree-level contributions. 

We have demonstrated the local matching of IR and threshold singularities between phase-space residues for both final and initial-state configurations. In the latter, the local cancellation of collinear singularities, although constrained by kinematics, leads to a local expression of collinear splitting functions. Explicit implementations of LTD causal unitary for benchmark decay rates with a detailed derivation of the contributing ingredients are presented in~\cite{LTD:2024yrb}.


\begin{acknowledgments}
We are deeply indebted to Stefano Catani, who was always a source of inspiration for the LTD. GR is warmly grateful to the staff of the European Research Council Executive Agency and the European Commission for their hospitality during the time this work was conceived. This work is supported by the Spanish Government (Agencia Estatal de Investigaci\'on MCIN/AEI/10.13039/501100011033) Grants No.~PID2020-114473GB-I00 and No.~PID2022-141910NB-I00, and Generalitat Valenciana Grant No. PROMETEO/2021/071. SRU is funded by CONAHCyT Grant No. 320856 (Paradigmas y Controversias de la Ciencia 2022), Ciencia de Frontera 2021-2042 and Sistema Nacional de Investigadores. PKD is supported by European Commission MSCA Action COLLINEAR-FRACTURE, Grant Agreement No.~101108573. GS is partially supported by H2020 STRONG-2020 Grant Agreement No. 824093 and H2020-MSCA-COFUND USAL4EXCELLENCE-PROOPI-391 Grant Agreement No 101034371.
\end{acknowledgments}

\section{Supplemental Material}

\appendix

\section{Local cancellation of quadratic and higher-order collinear singularities}

Vacuum diagrams with self-energy insertions involve Feynman propagators raised to a certain power. In general, Feynman diagrams with raised propagators will be the norm at very high perturbative orders. The corresponding LTD representation leads to causal propagators which are also raised to the same powers. In the core text of this Letter, we provided a simplified proof of how collinear singularities cancel between phase-space residues representing final states with different numbers of particles, assuming linear singularities. In this Supplemental Material attachment, we extend that proof to the cancellation of quadratic and higher-order singularities, along with the associated subleading singularities. For simplicity, we assume a scalar vacuum amplitude. This is sufficient to clearly illustrate the mechanism of the local cancellation of higher-order singularities. We refer the interested reader to Ref.~\cite{LTD:2024yrb} for the explicit expressions arising in physical decay rates, where the cancellation of quadratic singularities is tested in full detail.

\begin{figure}[t]
\begin{center}
\includegraphics[scale=.6]{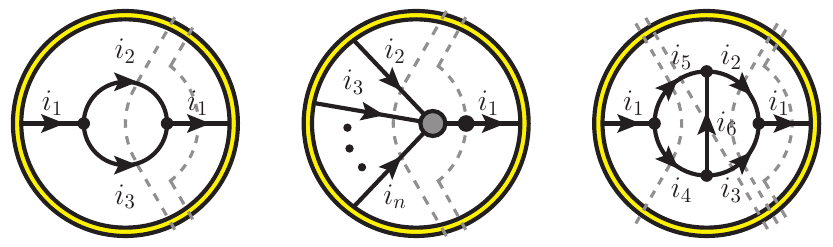}

{a) \hspace{4.5cm} b) \hspace{4.5cm} c)}

\caption{Diagrammatic interpretation of the local cancellation of higher-order singularities due to raised propagators. Each phase-space residue is identified with the interference of scattering amplitudes with different numbers of external particles, namely different numbers of loops. The two propagators labeled $i_1$ are identical and go on shell simultaneously. The dot over $i_1$ in the middle diagram represents an arbitrary power. 
\label{fig:finalstateself}}
\end{center}
\end{figure}

We consider first the case of quadratic singularities due to self-energy insertions at one loop (see Fig.~\ref{fig:finalstateself}a), where the propagator representing the particle labeled $i_1$ is squared. The kernel dual vacuum amplitude contains terms which are of the form 
\beq
\ad{\Lambda} \sim \frac{1}{\lambda_{i_3 i_2 i_X} \lambda_{i_1} \lambda_{i_1 i_X}}
\left( \frac{1}{\lambda_{i_1}}+\frac{1}{\lambda_{i_1 i_X}} \right)~.
\label{eq:doslambdasself}
\eeq
This expression was obtained with the package {\tt Lotty}~\cite{TorresBobadilla:2021dkq}. The index $i_X$ represents a set of particles common to the causal propagators appearing in \Eq{eq:doslambdasself}, i.e. $\lambda_{i_1 i_X} = \qon{i_1} + \sum_{i_s\in X} \qon{i_s} $ and $\lambda_{i_3 i_2 i_X} = \qon{i_3} + \qon{i_2} + \sum_{i_s\in X} \qon{i_s} $, with $\lambda_{i_1}=\qon{i_1}$. The corresponding phase-space residues are
\bea
\ad{\Lambda}(i_1 i_X) &=& {\rm Res} \left( \ad{\Lambda}, \lambda_{i_1 i_X} \right) \nn \\ &\sim&
\frac{1}{\lambda_{i_3 i_2 \bar i_1}\lambda_{i_1}}
\left( \frac{1}{\lambda_{i_1}} - \frac{1}{\lambda_{i_3 i_2 \bar i_1}} \right)~,
\label{eq:res1}
\eea
and
\bea
\ad{\Lambda}(i_3 i_2 i_X) &=& {\rm Res} \left( \ad{\Lambda}, \lambda_{i_3 i_2 i_X}\right) \nn \\ &\sim&
- \frac{1}{\lambda_{i_3 i_2 \bar i_1} \lambda_{i_1}}
\left( \frac{1}{\lambda_{i_1}} - \frac{1}{\lambda_{i_3 i_2 \bar i_1}} \right)~.
\label{eq:res2}
\eea
Both \Eq{eq:res1} and \Eq{eq:res2} are quadratically singular at $\lambda_{i_3 i_2 \bar i_1}\to 0$, and also exhibit a subleading linear singularity. However, the sum of their contributions to the cross section is finite,
\bea
&& \lim_{\lambda_{i_3i_2 \bar i_1} \to 0}
\left( \ad{\Lambda}(i_1 i_X)\, \ps{i_1 i_X} \right. \nn \\ && \qquad + \left. 
\ad{\Lambda}(i_3 i_2 i_X)\, \ps{i_3 i_2 i_X} \right) = {\cal O} \left( \lambda_{i_3 i_2 \bar i_1}^0\right)~,
\eea
where the $\ps{}$ functions encode energy conservation for each term. 

The generalization to configurations with an arbitrary number of particles (see Fig.~\ref{fig:finalstateself}b), where one of the propagators is raised to the power of $a_1$, is as follows 
\beq
\ad{\Lambda} \sim \frac{1}{\lambda_{i_n \cdots i_2 i_X}} \sum_{k=1}^{a_1} \frac{c_k}{\left(\lambda_{i_1}\right)^{2a_1-k} \left(\lambda_{i_1 i_X} \right)^{k}}~.
\label{eq:anylambdasself}
\eeq
The \Eq{eq:anylambdasself} also fulfils a recursive identity, where, according to Ref.~\cite{Aguilera-Verdugo:2020kzc}, the expression for a power of $a_1$ is the derivative of the expression for a power of $a_1-1$:
\beq
\left. \ad{\Lambda} \right|_{a_1} = \frac{1}{(a_1-1)} \frac{\partial}{\partial \lambda_{i_1}^2} \left. \ad{\Lambda} \right|_{a_1-1}~.
\eeq
The coefficients $c_k$ depend on the power $a_1$, although their exact values are not relevant to the discussion. For nonscalar vacuum amplitudes, they will generally be a rational function in the on-shell energies without poles in $\lambda_{i_1 i_X}$ or $\lambda_{i_n \cdots i_2 i_X}$. Without loss of generality, we can assume that~$c_k$ is a polynomial in~$\lambda_{i_X}$, i.e. $c_k(\lambda_{i_X})$. The detailed form of the tree or loop interaction represented by the gray blob in Fig.~\ref{fig:finalstateself}b is also not relevant.

The phase-space residues of \Eq{eq:anylambdasself} are
\bea
&& \ad{\Lambda}(i_1 i_X) = {\rm Res} \left( \ad{\Lambda}, \lambda_{i_1 i_X}\right) \nn \\ && \qquad \sim \frac{1}{\lambda_{i_n \cdots i_2 \bar i_1}} \sum_{k=1}^{a_1} \frac{(-1)^{k-1} c_k (-\lambda_{i_1})}{\left(\lambda_{i_1}\right)^{2a_1-k} \left(\lambda_{i_n \cdots i_2 \bar i_1} \right)^{k}}~,
\label{eq:res3}
\eea
and
\bea
&& \ad{\Lambda}(i_n \dots i_2 i_X) = {\rm Res} \left( \ad{\Lambda}, \lambda_{i_n \cdots i_2 i_X}\right) \nn \\ && \qquad \sim \frac{1}{\lambda_{i_n \cdots i_2 \bar i_1}} \sum_{k=1}^{a_1} \frac{(-1)^{k} c_k(-\lambda_{i_n\cdots i_2})}{\left(\lambda_{i_1}\right)^{2a_1-k} \left(\lambda_{i_n \cdots i_2 \bar i_1} \right)^{k}}~.
\label{eq:res4}
\eea
As before, there is a local cancellation of singularities for all orders in $\lambda_{i_n\cdots i_2 \bar i_1}\to 0$, 
\bea
&& \lim_{\lambda_{i_n \cdots i_2 \bar i_1} \to 0}
\left( \ad{\Lambda}(i_1 i_X) \,\ps{i_1 i_X} \right. \nn \\ && \left. + 
\ad{\Lambda}(i_n \cdots i_2 i_X)\, \ps{i_n \cdots i_2 i_X} \right) = {\cal O} \left( \lambda_{i_n \cdots i_2 \bar i_1}^0\right)~.
\eea

To conclude this section, we analyze in detail the case of a two-loop self-energy insertion. The configuration depicted in Fig.~\ref{fig:finalstateself}c is the most complex one contributing to a two-loop self-energy insertion, as it involves six Feynman propagators, or equivalently four entangled causal propagators. The vacuum amplitude has a term of the form 
\beq
\ad{\Lambda} \sim \frac{1}{\lambda_{i_5 i_4 i_X} \lambda_{i_6 i_5 i_3 i_X} \lambda_{i_3 i_2 i_X} \lambda_{i_1} \lambda_{i_1 i_X}} 
\left( \frac{1}{\lambda_{i_1}}+\frac{1}{\lambda_{i_1 i_X}} \right)~.
\label{eq:fourlambdasself}
\eeq
There are four phase-space residues encoded by \Eq{eq:fourlambdasself}: 
\bea
&& {\rm Res} \left( \ad{\Lambda}, \lambda_{i_3 i_2 i_X}\right) \nn \\ && \quad \sim  \frac{1}{\lambda_{i_6i_5\bar i_2} \lambda_{i_2 i_3\bar i_4\bar i_5} \lambda_{i_3 i_2\bar i_1} \lambda_{i_1}} \left( \frac{1}{\lambda_{i_1}} - \frac{1}{\lambda_{i_3 i_2\bar i_1}}\right)~, \nn \\
&& {\rm Res} \left( \ad{\Lambda}, \lambda_{i_6 i_5 i_3 i_X}\right) \nn \\ && \quad \sim \frac{1}{\lambda_{i_6i_5\bar i_2} \lambda_{i_6 i_3\bar i_4} \lambda_{i_1} \lambda_{i_6 i_5 i_3 \bar i_1}} \left( - \frac{1}{\lambda_{i_1}} + \frac{1}{\lambda_{i_6 i_5 i_3 \bar i_1}}\right)~, \nn \\
&& {\rm Res} \left( \ad{\Lambda}, \lambda_{i_5 i_4 i_X}\right) \nn \\ && \quad \sim  \frac{1}{\lambda_{i_2i_3\bar i_4 \bar i_5} \lambda_{i_6 i_3\bar i_4} \lambda_{i_1} \lambda_{i_4 i_5\bar i_1}} \left( - \frac{1}{\lambda_{i_1}} + \frac{1}{\lambda_{i_4 i_5\bar i_1}} \right)~,
\eea
and ${\rm Res} \left( \ad{\Lambda}, \lambda_{i_1 i_X}\right)$ is such that it cancels the singularities of the other three phase-space residues. 

If we now reverse the propagation direction of $i_4$ in Fig.~\ref{fig:finalstateself}c, this configuration is encoded by a term similar to \Eq{eq:fourlambdasself} with the substitution $\lambda_{i_5 i_4 i_X}\to \lambda_{i_6 i_4 i_3}$. For this configuration, the local cancellation of singularities occurs between the three phase-space residues involving $i_X$. The proof of the local cancellation of singularities for the other configurations allowed by causality proceeds in a similar way. Also at two loops we have to consider configurations where two Feynman propagators are raised to a power. The proof is more tricky, since it concerns terms with a nontrivial combination of raised causal propagators, but follows the same algorithmic procedure.

\section{Vacuum amplitude of a benchmark process}
\label{sec:vacuum}

\begin{figure}[t]
\begin{center}
\includegraphics[scale=.7]{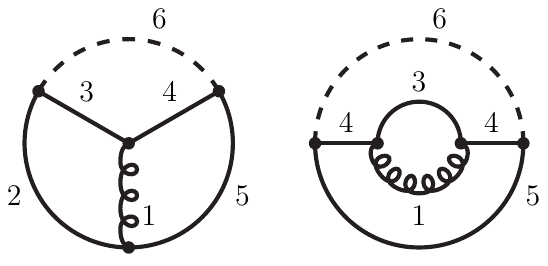}
\caption{Representative three-loop vacuum diagrams contributing to the Higgs decay into a quark-antiquark pair at NLO. The particle labeled~$1$ is a gluon, the particles~$2$ to~$5$ are quarks and the particle~$6$ represents a Higgs boson. 
\label{fig:higgsqqbar}}
\end{center}
\end{figure}

For the convenience of the reader, we present in this section an explicit expression for the three-loop vacuum amplitude in LTD from which the decay of a Higgs boson into a quark-antiquark pair at NLO is obtained following the method described in the core of the Letter. The representative vacuum diagrams are shown in Fig.~\ref{fig:higgsqqbar}.  The corresponding LTD vacuum amplitude is given by 
\bea
&& \ad{3}(Hq\bar q g) = \frac{4 g_H^{(3)}}{x_{123456}} \Bigg\{ \lambda_6^4 (1+\beta^2) \beta^2 \nn \\ && \times \left[ \left( \frac{1}{\lambda_{134}} + \frac{1}{\lambda_{456}} \right) \left( \frac{1}{\lambda_{125}} + \frac{1}{\lambda_{236}} \right) \frac{1}{2\lambda_{1356}} +
\frac{1}{\lambda_{134} \lambda_{456} \lambda_{2345}} \right] \nn \\
&& + \frac{\lambda_2}{\lambda_4} \Bigg[ 
\frac{\lambda_6^2 \beta^2}{\lambda_{456}}
\left( (d-2) \lambda_{1\bar3}
\left( \frac{1}{\lambda_4} + \frac{1}{\lambda_{456}}\right) - \frac{2m^2}{\lambda_{456}} \left( \frac{1}{\lambda_{134}} + \frac{1}{\lambda_{1356}}\right) \right)\nn \\
&& - \frac{2m^2}{\lambda_{134}} \Bigg( \frac{\lambda_6^2 \beta^2}{\lambda_{134}}
\left( \frac{1}{\lambda_{456}} + \frac{1}{\lambda_{1356}} \right) + 2\lambda_{5\bar 6}
\left( \frac{1}{\lambda_4} + \frac{1}{\lambda_{134}} \right) \Bigg) \Bigg] \nn \\ &&
+ \lambda_1 \lambda_6^2 \left( \frac{1}{2\lambda_{236}} + \frac{1}{\lambda_{2345}} \right) \frac{(d-2)\beta^2 - 2}{\lambda_{456}}
\nn \\ 
&& + \lambda_2 \bigg[\left( \left( 2\lambda_6^2 \beta^2 \left(2-\frac{m^2}{\lambda_4^2} \right) - (d-2)\lambda_{236} \lambda_{\bar 2 36} \right) \right. \nn \\ && \left. \times \left(\frac{1}{\lambda_{134}} + \frac{1}{\lambda_{456}} \right) + 2(d-2) \lambda_4 \right) \frac{1}{\lambda_{1356}} \nn \\
&& + \left( 2\lambda_6^2 \beta^2 \left(2-\frac{m^2}{\lambda_4^2} \right) - (d-2)\lambda_{2\bar 3 6} \lambda_{\bar 2 \bar 3 6} \right) \frac{1}{\lambda_{134} \lambda_{456}} \nn \\ && + \frac{(d-2)\lambda_{1\bar 3}\lambda_{5\bar 6}}{\lambda_4^2} \bigg] \nn \\
&& -2 \lambda_6 \Bigg[\left( \frac{1}{2\lambda_{125}} + \frac{1}{\lambda_{2345}} \right) \frac{2\lambda_6^2-6m^2+\lambda_{456}\lambda_{45\bar 6}}{\lambda_{134}} \nn \\ && 
- \frac{2\lambda_4\lambda_5}{\lambda_{134}\lambda_{125}}  \Bigg] + (d-4) \Bigg[ \frac{\lambda_6 \lambda_{25}-2\lambda_2\lambda_5}{\lambda_{134}} - \frac{\lambda_1 \lambda_6}{\lambda_{2345}} \Bigg] \Bigg\} \nn \\
&& + (2\leftrightarrow 3, 4\leftrightarrow 5) + (2\leftrightarrow 4, 3\leftrightarrow 5)+ (2\leftrightarrow 5, 3\leftrightarrow 4)~,
\label{eq:vacuum}
\eea
where 
\beq
x_{123456} = \prod_{s=1}^6 2\qon{s}~, 
\qquad \beta = \sqrt{1-4m^2/\lambda_6^2}~,
\eeq
with $m$ the mass of the quark. 
The factor $g_H^{(3)}$ encodes the Yukawa and strong couplings, and the color factor, $g_H^{(3)} = y_q^2 \gs^2 C_F$. The corresponding phase-space residues are straightforwardly obtained from \Eq{eq:vacuum} as the residues for $\lambda_{236} \to 0$ and $\lambda_{456} \to 0$, which represent the interference of the one-loop amplitude with the tree-level, and for $\lambda_{1246} \to 0$ and $\lambda_{1356} \to 0$, which represent the squared amplitudes with an extra gluon radiation. Explicit expressions are presented in Ref.~\cite{LTD:2024yrb}.

\bibliography{LTD_causal_unitary_fundamentals}

\end{document}